\documentclass[aps,prx,showpacs,
amsmath,amssymb,longbibliography,showkeys,12pt]{revtex4-2}%
\usepackage{amsfonts}
\usepackage{amsmath}
\usepackage{amssymb}
\usepackage{graphicx}
\usepackage{epsfig}
\usepackage{exscale}
\usepackage{float}
\usepackage{bm}
\usepackage{suffix}
\usepackage{mathtools}
\usepackage{newunicodechar}
\usepackage{lipsum}
\usepackage{enumitem}
\usepackage{bbding}
\usepackage{tikz}
\usepackage{multirow}
\usepackage{nameref}
\usepackage[colorlinks=true,linkcolor=blue]{hyperref}%
\providecommand{\U}[1]{\protect\rule{.1in}{.1in}}
\expandafter\ifx\csname package@font\endcsname\relax\else
\expandafter\expandafter
\expandafter
\expandafter
\expandafter{\csname package@font\endcsname}
\fi
\DeclarePairedDelimiterX\MeijerM[3]{\lparen}{\rparen}
{\begin{smallmatrix}#1 \\ #2\end{smallmatrix}\delimsize\vert\,#3}
\newcommand\MeijerG[8][]{  G^{\,#2,#3}_{#4,#5}\MeijerM[#1]{#6}{#7}{#8}}
\WithSuffix
\newcommand\MeijerG*
[7]{
G^{\,#1,#2}_{#3,#4}\MeijerM*{#5}{#6}{#7}}

\begin{document}
\title{Dependence of Critical Exponents Accuracy on the Number of Fields in the \texorpdfstring{$O(N)$}{TEXT}-Invariant \texorpdfstring{$\phi^4$}{TEXT} Model}
\author{Abouzeid M. Shalaby}
\affiliation{Department of Physics and Materials Sciences, College of Arts and Sciences, Qatar University, Al
Tarfa, Doha 2713, Qatar}
\begin{abstract}

The study of large-$N$ cases  of the $O(N)$-symmetric model at criticality has attracted significant attention in recent years. In particular, a recent high-precision Monte Carlo study ( Physical Review B 105, 054428 (2022)) reported the most accurate estimates to date for the critical exponents $\nu$, $\eta$, and $\omega$ for $N \ge4$. Within the framework of the renormalization group (RG)—the cornerstone of the modern theory of critical phenomena—powerful computational approaches such as the   $\varepsilon$-expansion and non-perturbative RG are available. Since the effective expansion parameter in the $\varepsilon$-expansion, $\sigma = \frac{3}{N+8}$, decreases with increasing N, one expects resummation techniques to become progressively more accurate in the large-N regime.
To test this expectation, we apply the entire-hypergeometric resummation algorithm developed by Shalaby et al. to the recently obtained seven-loop divergent   $\varepsilon$-series for $N \ge 4$, yielding high-precision estimates for $\nu$, $\eta$, and $\omega$ . Our analysis confirms the anticipated improvement in accuracy with increasing N. To assess the significance of these results, we note that at the same seven-loop order the $O(2)$ case exhibits errors that are an order of magnitude larger than those obtained from experiment, Monte Carlo simulations, and conformal-field-theory analysis. In contrast, for sufficiently large N, the errors in the present work are of the same order of magnitude as those from Monte Carlo and non-perturbative RG methods, demonstrating the strong predictive power of our resummation approach in the large-N regime.

\end{abstract}
\maketitle

\section{ Introduction}

The $O(N)$-vector model plays a fundamental role in the study of critical phenomena across field theory and condensed matter physics. It provides predictive insights into critical parameters that characterize various kinds of phase transitions, including quantum phase transitions as well as those observed in superfluidity and magnetism. For large $N$, the model has been extensively investigated, particularly in recent years. In three dimensions, the $O(N)$-symmetric scalar $\phi^4$ model provides an effective framework for analyzing critical behavior in a wide range of physical systems. In the context of our work, where $N\geq4$, the $O(4)$ scalar $\phi^4$ model describes the finite-temperature phase transition in QCD with two light flavors \cite{QCD}. Additionally, the $O(10)$ model is relevant for neutron stars, while the $O(18)$ model applies to the superfluid phase transition in ${}^3He$.

The phase transition within the $O\left(  N\right)$-vector model for large $N$ has recently exposed to different theoretical investigations using
different  approaches. In Ref.\cite{MCON4}, the so far most precise  prediction of the critical exponents have been obtained by Martin Hasenbusch using Monte Carlo simulations. The functional renormalization group has investigated the same model for large $N$ and obtained precise results too \cite{NPR2020}. The large-$N$ expansion, which is exact for sufficiently large-$N$ values, has been improved recently using   2PI calculations \cite{2PILN,2PI2011}. Apart from these methods, the perturbative renormalization group (RG) calculations have achieved recently a relatively high order in loop expansion of the renormalization group functions. In Refs.\cite{Schnetz2023,Schnetz-06-23,Schnetz2018}, the seven-loop order has been obtained for  different renormalization group functions of the $O\left(N\right)  $-vector model. For these expansions, one can extract the corresponding $\varepsilon$-expansion which is always more accurate than the original coupling-expansion. However, both series ( $\varepsilon$ and $g$ expansions) are asymptotic and do have zero-radius of convergence which means that they are useless by their own unless we treat them using a resummation technique ( for instance).

Borel resummation remains one of the most prominent and widely adopted techniques for assigning meaningful values to divergent perturbative expansions. In recent applications \cite{Borel-6L}, it has been supplemented by conformal mapping to resum the six-loop series of the $\varepsilon$-expansion. In a very recent work \cite{entire}, we proposed an alternative hypergeometric-resummation framework based on entire hypergeometric functions. For the same amount of perturbative input, this entire-hypergeometric approach is more direct in implementation while maintaining a high level of accuracy when compared with results obtained from more sophisticated methods, as illustrated in Table~\ref{N4EX}. An additional advantage of the method is that it goes beyond the numerical estimation of critical exponents: it can also infer nonperturbative information, including strong-coupling behavior and large-order growth parameters, from the finite set of perturbative coefficients available in the original series.

 Because the entire-hypergeometric resummation can determine the large-order growth parameters of a perturbative series, it can also provide insight into the nature of the singularity closest to the origin of the Borel series, namely whether it is associated with a renormalon or an instanton contribution. The series considered in the present work was recently analyzed from this perspective in Ref.~\cite{Renormalon22}, where it was found to exhibit a tendency consistent with an instanton singularity rather than a renormalon one. This outcome is physically well motivated.  The $\varepsilon$-expansion is organized around the upper critical dimension $d=4$, where the $\phi^{4}$ theory is renormalizable and the zeroth-order term reproduces the exact mean-field exponents. Corrections in powers of $\varepsilon=4-d$ then measure the effect of fluctuations below four dimensions. In three and two dimensions, the theory becomes super-renormalizable, so the ultraviolet structure is much softer and the mechanism usually associated with renormalon singularities is not expected to dominate \cite{Ren-2005,ren-rep}. The leading large-order behavior is therefore naturally attributed to instanton-type saddle points, in agreement with the tendency observed in the present series. 

The entire-hypergeometric algorithm is able to resum a divergent series of Gevrey$-m$ $(m\geq1)$ type where it has a large-order behavior of their
coefficients that takes the form:
\begin{equation}
c_{i}\sim\left(  mi\right)!\ i^{b}\sigma^{i},\text{ for large }i\text{,}%
\end{equation}
assuming  for a physical amplitude $Q\left(  x\right)$ we have the divergent series:
\begin{equation}
Q\left(  x\right)  =\sum_{i=0}^{\infty}c_{i}x^{i},
\end{equation}
with $x$ is the perturbation parameter. For the critical exponents within the $O\left(  N\right)  $ model $x$ will be represented by $\varepsilon=4-D$ where $D$ is the dimension of the Euclidean space-time.

The algorithm start by assigning a hypergeometric function ${}_pF_{q}$ where $p=q+m+1$ to represent the original series by matching the coefficients of the original series to the corresponding ones of the ${}_pF_{q}$ hypergeometric series. This step defines the numerator $\left\{  a_{l}\right\}  $   and
denominator $\left\{  b_{j}\right\}  $ parameters of the hypergeometric function ${}_pF_{q}(a_{1},...a_{p};b_{1}....b_{q};\sigma z)$. Note that the condition $p=q+m+1$ guarantees that ${}_pF_{q}$ to have the same form of the large-order asymptotic behavior as the original series.

The series of interest here in this work is   a Gevery-$1$ type and thus the suitable candidate from hypergeometric functions is ${}_pF_{p-2}$. For instance, if we  have the first $M$ orders from the divergent series $\sum_{i=0}^{M}c_{i}x^{i}$, then we mach them by the first $M$ orders from  the ${}_pF_{p-2}$ hypergeometric series. The analytic continuation is then obtained via the integral representation of the hypergeometric function which is a Mellin--Barnes integral such that\ \cite{Analytic2016}:
\begin{equation}
_{\text{ }p}F_{q}(a_{1},...a_{p};b_{1}....b_{q};z)=\frac{\prod_{k=1}^{q}%
\Gamma\left(  b_{k}\right)  }{\prod_{j=1}^{p}\Gamma\left(  a_{j}\right)
}\frac{1}{2\pi i}\int_{C}\frac{\ \Gamma\left(  s\ \right)  \prod_{j=1}%
^{p}\Gamma\left(  a_{j}-s\right)  }{\ \prod_{k=\ 1}^{q}\Gamma\left(
b_{k}-s\ \right)  }\left(  -z\right)  ^{-s}ds, \label{hyp-G-C2}%
\end{equation}
 which in turn can be represented as:
\begin{align}
&  \,_{\text{ }p}F_{q}(a_{1},...a_{p};b_{1}....b_{q};\sigma\text{
}z)\nonumber\\
&  =\frac{\prod_{i=1}^{q}\Gamma\left(  b_{i}\right)  }{\prod_{i=1}^{p}%
\Gamma\left(  a_{i}\right)  }\sum_{k=1}^{p}(-\sigma\text{ }z)^{-a_{k}}%
\frac{\Gamma\left(  a_{k}\right)  }{\prod_{j=1}^{q}\Gamma\left(  b_{j}%
-a_{k}\right)  }\prod_{j=1,j\neq k}^{p}\Gamma\left(  a_{j}-a_{k}\right)
\times\nonumber\\
&  \ \,_{q+1}F_{p-1}\left(  a_{k},a_{k}-b_{1}+1,.....,a_{k}-b_{q}%
+1;-a_{1}+a_{k}+1,\underbrace{.....}_{\ast},-a_{p}+a_{k}+1;\frac{(-1)^{p-q+1}%
}{\sigma\text{ }z}\right)  . \label{entire}%
\end{align}
The asterisk here means that $p\neq k$. Note that the left-hand side is a divergent series while for $p\geq q+2$, the right-hand side is a sum over
entire-hypergeometric functions. This means that while the input information are just the coefficients in a given  divergent series, the output is represented in terms of convergent ones.

\begin{figure}[H]
\begin{center}
\includegraphics[width=10cm,height=7cm]{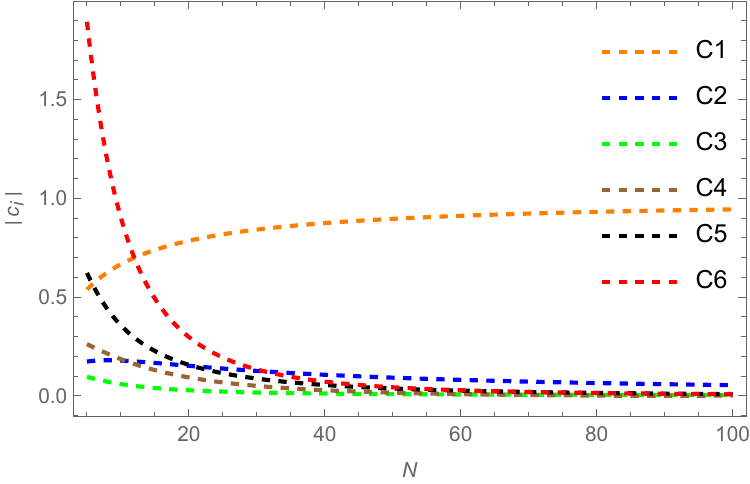}
\end{center}
\caption{\textit{The   first six coefficients in the series for $\nu^{-1}$ in Eq.(\ref{7Lexponenets}) plotted versus $N$. One can easily realize that except for $c_1$, the other coefficients ( including $c_7$ which is not appearing in the figure) are decreasing functions in $N$. Also, note that the decrease is sharper for larger $N$ values.}}
\label{CNvN}
\end{figure}

For the renormalization group functions in the $O(N)-$vector model when represented in the $\varepsilon-$ expansion form, we realized that the coefficients of the different series are all  including the parameter $\sigma=\frac{3}{N+8}$. This means that for large$-N$ ( which are the cases of interest for this work), the coefficients are getting smaller   and in fact all orders higher than the first order  tend to zero as $N\rightarrow\infty$ ( see Fig.\ref{CNvN}) giving the expected exact value predicted from the large-$N$ expansion. This led us to expect that our resummation method will give highly accurate results for the cases of interest $(N\geq4)$. In fact, the best resummation result for O(2) case has an error that is one order of magnitude higher than the ones for experiment, MC and CB results. This makes the RG results are not   able to share in the well-known $\lambda$ point dispute of the super-fluid phase transition in ${}^4He$ ( See our previous work in Ref.\cite{lambda} and the references therein).  As will be shown later in this work, as $N$ increases, the uncertainty in the resummed series becomes comparable in order of magnitude to those reported in Monte Carlo and NPRG calculations.

The error  in the calculation  can be determined via well known methods \cite{lambda}. For the algorithm we use, there are two main sources of errors. The first one is due to the so far unknown higher orders  in the perturbation series used while the other source comes from the arbitrary
parameters used in the resummation algorithm where due to reasons that have been explained in our previous work \cite{lambda} we will take it as the large-order parameter $\sigma.$ We shall variate $\sigma$ around its exact known value for the $O\left(  N\right)  $ model \cite{Kleinert-Borel} which takes the form $\sigma=\frac{3}{N+8}$. In varying $\sigma$ we seek a less sensitive region (plateau) to the variations.  Keeping this in mind and  as explained in Ref.\cite{lambda} our error will take the shape:
\begin{equation}
\Delta\left(  \sigma,k\right)  =\left\vert \nu^{7}\left(  \sigma\right)
-\nu^{6}\left(  \sigma\right)  \right\vert +Var_{\sigma}\left(  \nu_{k}%
^{7}\left(  \sigma\right)  \right)  ,
\end{equation}
where
\[
Var_{\sigma}\left(  \nu_{k}^{7}\left(  \sigma\right)  \right)  =\min
_{x\leq\sigma\leq y}\left(  \max_{x\leq\sigma^{\prime}\leq y}\left(  \nu
_{k}^{7}\left(  \sigma\right)  -\nu_{\acute{k}}^{7}\left(  \sigma^{\prime
}\right)  \right)  \right)  .
\]
Here $6$ and $7$ are  labeling the six and seven loops estimates ( for the error in $7$-loop calculations). 

The structure of the paper is as follows. In Sec.~\ref{ON}, we introduce the renormalization-group functions of the $O(N)$ model and outline the procedure used to obtain the $\varepsilon$-expansion from the loop expansion. In Sec.~\ref{ON4}, we present the critical-exponent estimates for the $O(4)$ case, using the seven-loop $\varepsilon$-expansion as input to the entire-hypergeometric algorithm. The corresponding results for the $O(5)$ model are reported in Sec.~\ref{ON5}, while the $O(10)$ case is discussed in Sec.~\ref{ON10}. 

For relatively large values of $N$, the higher-order coefficients of the series, $c_i$ with $i\geq 2$, become smaller, suggesting that more precise resummed estimates may be obtained. We therefore also consider the cases $N=20$ and $N=100$; the corresponding results are presented in Sec.~\ref{ON20100}. Finally, a summary and concluding remarks are given in Sec.~\ref{conc}.

\section{The seven-loop renormalization group functions in the \texorpdfstring{O$\left(  N\right)$}{TEXT} vector model}\label{ON}

We  stress the critical exponents within  the $O\left(  N\right)  -$vector model where it has the Lagrangian density of the
from: 
\[
\mathcal{L=}\frac{1}{2}\left(  \partial\Phi\right)  ^{2}+\frac{m^{2}}{2}%
\Phi^{2}+\frac{\lambda}{4!}\Phi^{4}.
\]
Here $\Phi=\left(  \phi_{1},\phi_{2},\phi_{3},...........\phi_{N}\right)  $ is an N-component field that have an $O(N)$ symmetry in the sense that $\Phi
^{4}=\left(  \phi_{1}^{2}+\phi_{2}^{2}+\phi_{3}^{2}+...........\phi_{N} ^{2}\right)  ^{2}$. Within the minimal-subtraction $\overline{MS}$ technique,
the seven-loop ( $g$-expansion, $g$ is the renormalized coupling) perturbation series for different renormalization group functions have been obtained in
Refs.\cite{Schnetz2018,Schnetz-06-23,Schnetz2023}. It is well known that the $\varepsilon-$expansion ought to give more accurate results than working with
$g$-series. At the critical point, the $\beta-$function is certainly zero which defines the critical coupling $g_c$. The equation for $g_c$ can lead to an expansion in terms of the parameter $\varepsilon$. This expansion can be substituted back in the other renormalization group functions $\gamma_{m}$ ( the mass anomalous dimension)  and $\gamma$ ( the anomalous dimension of the scalar field) to get the up to $O\left(  \varepsilon^{7}\right)  $ series. These series are divergent as well as defining the critical exponents $\nu$ and $\eta$. For the approach to scaling exponent $\omega$, one can get its series from the relation $\omega =\beta^{^{\prime}}\left(  g_{c}\right)  $ where $g_{c}$ is the critical coupling. Keeping these in mind,  we can obtain the flowing relations \cite{Truncation}:%
\begin{align}\label{7Lexponenets}%
\nu^{-1}  &  =2+\gamma_{m}\left(  g_{c}\right)  =2-\frac{\varepsilon\left(
N+2\right)  }{\left(  N+8\right)  }-\ \frac{\varepsilon^{2}\left(  N+2\right)
}{2\left(  N+8\right)  ^{3}}\left(  13N+44\right)  +..........+O\left(
\varepsilon^{7}\right)  ,\nonumber\\
\eta &  =\ \frac{\varepsilon^{2}\left(  N+2\right)  }{2\left(  N+8\right)
^{2}}+\frac{\varepsilon^{3}\left(  N+2\right)  }{8\left(  N+8\right)  ^{4}%
}\left(  -N^{2}+56N+272\right)  +..........+O\left(  \varepsilon^{7}\right)
,\nonumber\\
\omega &  =\varepsilon-\ \frac{3\varepsilon^{2}\ }{\left(  N+8\right)  ^{2}%
}\left(  3N+14\right)  +..........+O\left(  \varepsilon^{7}\right)  .
\end{align}
All of these series are divergent as they have a large-order behavior of the form $i!n^{b}\sigma^{i}$.  Accordingly, they can be approximated by a
hypergeometric function $_{\text{ }p}F_{p-2}(a_{1},...a_{p};b_{1}....b_{p-2};\sigma z)$ which in turn can be analytically continued using the representation in Eq.(\ref{entire}). In the following sections, we list the results of our entire-hyper geometric approximation for the critical exponents $\nu,\eta$ and $\omega$ for different Large-$N$ values.  
\section{Critical exponents for the \texorpdfstring{O$\left(  4\right)$}{TEXT} vector model}\label{ON4}

The $N=4$ case can describe the critical behavior in QCD\ with two light flavors. For that case we have $\sigma=\frac{1}{4}$ while the seven-loop series are then given by: 
\begin{align}
\nu^{-1}  &  \simeq2.00000-0.500000\varepsilon-0.166667\varepsilon
^{2}+0.105856\varepsilon^{3}-0.278661\varepsilon^{4}\nonumber\\
&  +0.702167\varepsilon^{5}-2.23369\varepsilon^{6}+7.97005\varepsilon
^{7}, \label{nu7}%
\end{align}

\begin{align}
\eta &  \simeq\ 0.0208333\varepsilon^{2}+0.0173611\varepsilon^{3}%
-0.00708518\varepsilon^{4}+0.0176315\varepsilon^{5}\nonumber\\
&  -0.0473628\varepsilon^{6}+0.152188\varepsilon^{7}, \label{eta7}%
\end{align}%
\begin{align}
\omega &  \simeq\varepsilon-0.541667\varepsilon^{2}+1.15259\varepsilon
^{3}-3.27193\varepsilon^{4}+10.8016\varepsilon^{5}\nonumber\\
&  -40.5665\varepsilon^{6}+166.256\varepsilon^{7}. \label{omega7}%
\end{align}

For the inverse exponent $\nu^{-1}$, the seven-loop approximation is
\[
\nu_{7}^{-1}=2.00000-0.500000\varepsilon_{\text{ }4}F_{2}(a_{1},a_{2}%
,a_{3},a_{4};b_{1},b_{2};\sigma\varepsilon),
\]

with the analytic continuation setp goes through the relation:%
\begin{align}
&  {}_4F_{2}(a_{1},a_{2},a_{3},a_{4};b_{1},b_{2};\sigma
\varepsilon)\nonumber\\
&  =\frac{\prod_{i=1}^{2}\Gamma\left(  b_{i}\right)  }{\prod_{i=1}^{4}%
\Gamma\left(  a_{i}\right)  }\sum_{k=1}^{4}(-\sigma\text{ }z)^{-a_{k}}%
\frac{\Gamma\left(  a_{k}\right)  }{\prod_{j=1}^{2}\Gamma\left(  b_{j}%
-a_{k}\right)  }\prod_{j=1,j\neq k}^{4}\Gamma\left(  a_{j}-a_{k}\right)
\times\nonumber\\
&  \ \,_{3}F_{3}\left(  a_{k},a_{k}-b_{1}+1,.....,a_{k}-b_{q}%
+1;-a_{1}+a_{k}+1,\underbrace{.....}_{\ast},-a_{4}+a_{k}+1;\frac{(-1)^{3}%
}{\sigma\text{ }z}\right)  .
\end{align}
The six-loop approximation is also needed for the error calculation which can
be approximated by:%
\begin{equation}
\nu_{6}^{-1}=2.00000 \ { }_4F_{2}(a_{1},a_{2},a_{3},a_{4};b_{1}%
,b_{2};\sigma\varepsilon).
\end{equation}
Note that both  $\nu_{7}^{-1} $ and $\nu_{6}^{-1}$ can be represented in a
different way by approximating the subtracted series. For instance the six-loop
can also be represented by:%
\begin{equation}
\nu_{6}^{-1}=2.00000-0.500000\varepsilon-0.166667\varepsilon^{2}\ { }_3
F_{1}(a_{1},a_{2},a_{3}\ ;b_{1}\ ;\sigma\varepsilon).
\end{equation}
Such arbitrariness has been taken into account in error calculation by taking  the representation
that minimizes the error.

Our algorithm predicts the value $\nu=0.7444(67)$. To test the precision of our calculation we mention that the recent Monte Carlo (MC) calculations give
the result  $\nu=0.74817(20)$ \cite{MCON4} while the conformal bootstrap (CB) calculations give the result $\nu=0.7508(34)$ \cite{CB2016}.

The series for $\eta$ exponent in Eq.(\ref{eta7}) can be rewritten as
\begin{align}
\eta_{7}+\varepsilon &  \simeq\varepsilon+\ 0.0208333\varepsilon
^{2}+0.0173611\varepsilon^{3}-0.00708518\varepsilon^{4}+0.0176315\varepsilon
^{5}\nonumber\\
&  -0.0473628\varepsilon^{6}+0.152188\varepsilon^{7}%
\end{align}
 which can have the approximation
\[
\eta_{7}+\varepsilon\simeq\varepsilon_{\text{ }4}F_{2}(a_{1},a_{2},a_{3}%
,a_{4};b_{1},b_{2};\sigma\varepsilon).
\]
For that series our algorithm gives the result $\eta=0.0363(10)$ compared to
MC result of $0.03624(8)$ \cite{MCON4} and the CB result $0.0378(32)$%
\cite{CB2015} while the non perturbative renormalization group (NPRG) result is $0.0360(12)$\cite{NPR2020}.

The approach to scaling critical exponent can also approximated by the hypergeometric function
\begin{equation}
\omega_{7}\simeq\varepsilon_{\text{ }4}F_{2}(a_{1},a_{2},a_{3},a_{4}%
;b_{1},b_{2};\sigma\varepsilon).
\end{equation}
Our prediction for $\omega$ yield the result $0.7486(24)$ compared to the MC prediction of $0.755(5)$\cite{MCON4} and the CB result of $0.817(30)$
\cite{CB2016} while NPRG result is $0.761(12)$\cite{NPR2020}.

As we expected, for large-N, our resummation technique gives very precise results. In table-\ref{N4EX}, one can find our predictions listed and compared to    results  from other methods. One can realize that the relatively simple algorithm we use gives precise results that are competitive to the  recent and more sophisticated methods used for the same problem. We also listed our six-loop results in the table. Note that $MC$ applied to that case took 8.5 years of CPU time (single core of an AMD EPYC${}^{TM}$ 7351P CPU) \cite{MCON4}. It is worth noting that the uncertainty of the six-loop result is estimated using the corresponding five-loop prediction. However, at five-loop order, the series for $\eta/\varepsilon^{2}$, for example, contains only three non-trivial coefficients. As a result, the associated hypergeometric approximant involves only a very limited number of parameters. This restriction may therefore affect the precision of the error estimate and, consequently, the quoted accuracy of the six-loop calculation. 

\begin{table}[H]
\caption{\protect\scriptsize
Critical exponents $\nu$, $\eta$, and $\omega$ of the $O(4)$ vector model obtained by applying the entire-hypergeometric resummation algorithm to the six- and seven-loop $\varepsilon$-expansion series. The results are compared with conformal-bootstrap (CB) estimates \cite{CB2016,CB2015}, Monte Carlo (MC) simulations \cite{MCON4}, and NPRG calculations \cite{NPR2020}.}
\label{N4EX}
\begin{center}

\begin{tabular}{|l|l|l|l| }
\hline
\ \ \ \ \ \ \  Method\ \ & $\ \ \ \ \ \  \nu$ & $\ \ \ \ \ \ \eta$ & $\ \ \ \ \ \omega$   \\ \hline
\ \ \begin{tabular}[c]{@{}l@{}}  \ \ \ CB(2016,2015,2016)\\  \ \ \ MC(2022)\\ \ \ \ $\varepsilon^6$: (This work) \\    \ \ \ $\varepsilon^7$:(This work)\\ \ \ \ NPRG(2020)\\\end{tabular}\ \ & \begin{tabular}[c]{@{}l@{}}0.7508(34)\\  0.74817(20) \\ 0.7412(26)\\  0.7444(67)\\ 0.7478(9)\\ \end{tabular} & \begin{tabular}[c]{@{}l@{}}0.0378(32)\\0.03624(8)\\ 0.0385(31)\\  0.0363(10)\\ 0.0360(12)\\ \end{tabular} & \begin{tabular}[c]{@{}l@{}}0.817(30)
 \\ 0.755(5)\\  0.773(52)\\ 0.7486(24)\\0.761(12)\\ \end{tabular}  \\ \hline 

\end{tabular}%
\end{center}
\end{table}

\section{The \texorpdfstring{$\nu,\eta$ and $\omega$}{TEXT} Critical exponents for the \texorpdfstring{O$\left(
5\right)  $}{TEXT} case}\label{ON5}

For this case, we have the seven-loop series of the form:%

\begin{align}
\nu^{-1}  &  \simeq\ 2.00000-0.538462\varepsilon-0.173646\varepsilon
^{2}+0.0964117\varepsilon^{3}-0.262298\varepsilon^{4}\nonumber\\
&  +0.622526\varepsilon^{5}-1.89386\varepsilon^{6}+6.42964\varepsilon^{7},
\end{align}
which can be  approximated as:
\[
\nu^{-1}\simeq2.00000-0.538462\ \varepsilon \ {}_4F_{2}(a_{1},a_{2}%
,a_{3},a_{4};b_{1},b_{2};\sigma\varepsilon).
\]
Our prediction for $N=5$ is $\nu=0.780(5)$ while the recent MC result is
$0.7802(6)$ \cite{MCON4}   and NPRG gives the result $0.7797(9)$\cite{NPR2020}. Again taking into account the sophistication in the other methods, our algorithm predicts great results.

For the $\eta$ exponent, the sven-loop result is
\begin{align}
\eta &  \simeq\ 0.0207101\varepsilon^{2}+0.0161453\varepsilon^{3}%
-0.00680080\varepsilon^{4}+0.0151257\varepsilon^{5}-\nonumber\\
&  0.0395631\varepsilon^{6}+0.120232\varepsilon^{7}.
\end{align}
This order yields the result $\eta=0.034591(55)$. The recent NPRG result is  $0.0338(11)$ \cite{NPR2020} while the more recent MC result is $0.03397(9)$%
\cite{MCON4}. 

For $\omega$, the seven-loop series is given by:

\begin{align*}
\omega &  =\varepsilon+0.514793\varepsilon^{2}+1.04243\varepsilon
^{3}-2.85996\varepsilon^{4}+8.98927\varepsilon^{5}\\
&  -32.1663\varepsilon^{6}+125.419\varepsilon^{7}.
\end{align*}
Our seven-loop resummation result is $\omega=0.7554(21)$. The recent result from MC\ is $0.754(7)$ \cite{MCON4} while NPRG result is $0.760(18)$\cite{NPR2020}. The  results   are listed in table-\ref{N5EX}
\begin{table}[H]
\caption{{\protect\scriptsize
Critical exponents $\nu$, $\eta$, and $\omega$ of the $O(5)$ vector model obtained by applying the entire-hypergeometric resummation algorithm to the six- and seven-loop $\varepsilon$-expansion series. The results are compared with Monte Carlo (MC) simulations \cite{MCON4} and NPRG calculations \cite{NPR2020}.
}}
\label{N5EX}
\begin{center}

\begin{tabular}{|l|l|l|l| }
\hline
\ \ \ \ \ \ \  Method\ \ & $\ \ \ \ \ \  \nu$ & $\ \ \ \ \ \ \eta$ & $\ \ \ \ \ \omega$   \\ \hline
\ \ \begin{tabular}[c]{@{}l@{}}  \ \ \ MC(2022)\\ \ \ \ $\varepsilon^6$: (This work)  \\ \ \ \ $\varepsilon^7$: (This work)\\ \ \ \ NPRG(2020)\\\end{tabular}\ \ & \begin{tabular}[c]{@{}l@{}} 0.7802(6) \\ 0.7714(15)\\ 0.780(5)\\  0.7797(9)\\ \end{tabular} & \begin{tabular}[c]{@{}l@{}}0.03397(9)\\ 0.037(3)\\   0.034591(55)\\ 0.0338(11)\\ \end{tabular} & \begin{tabular}[c]{@{}l@{}} 0.754(7)\\ 0.773(31)\\  0.7554(21)\\0.760(18)\\ \end{tabular}  \\ \hline 

\end{tabular}%
\end{center}
\end{table}

\section{Critical exponents for the \texorpdfstring{O$\left(  10\right)$}{TEXT} case}\label{ON10}

This case also has been investigated recently using MC\ in Ref.\cite{MCON4} and NPRG in Ref.  \cite{NPR2020}.

For $N=10$, the seven-loop $\varepsilon$-expansion for the $\nu,\eta$ and $\omega$ critical exponents take the form:

\begin{align}
\nu^{-}  &  =2.00000-0.666667\varepsilon-0.179012\varepsilon^{2}%
+0.0597939\varepsilon^{3}-0.185638\varepsilon^{4}\nonumber\\
&  +0.356941\varepsilon^{5}-0.901457\varepsilon^{6}+2.5213\varepsilon^{7},
\end{align}

\begin{align*}
\eta &  =0.0185185\varepsilon^{2}+0.0104595\varepsilon^{3}%
-0.00595899\varepsilon^{4}+0.00678457\varepsilon^{5}\\
&  -0.0181265\varepsilon^{6}+0.0430667\varepsilon^{7},
\end{align*}

\begin{align*}
\omega &  =\varepsilon-0.407407\varepsilon^{2}+0.687784\varepsilon
^{3}-1.63261\varepsilon^{4}+4.26617\varepsilon^{5}\\
&  -12.5796\varepsilon^{6}+40.3282\varepsilon^{7}.
\end{align*}
Our seven-loop prediction for the critical exponent $\nu$ gives the result $0.8792(9)$ while the recent MC calculations yield the result $0.8797(9)$
\cite{MCON4} \ and the recent NPRG result is $0.8776(10)$\cite{NPR2020}.

For $\eta$, we obtained the result $0.02402(25)$. MC calculations in Ref.\cite{MCON4} gives the result $0.02302(12)$ and NPRG\ result is $0.0231(6)$ \cite{NPR2020}. Likewise we get the value $0.7894(18)$ for the critical exponents $\omega$ compared to the MC result $0.816(16)$ \cite{MCON4}
and NPRG result of $0.807(7)$\cite{NPR2020}. The results for $\nu,\eta$ and $\omega$  are listed in table-\ref{N10EX}.

\begin{table}[H]
\caption{\protect\scriptsize
Six- and seven-loop entire-hypergeometric resummation results for the critical exponents of the $O(10)$ vector model. For comparison, we include Monte Carlo (MC) simulations \cite{MCON4} and NPRG calculations \cite{NPR2020}.}
\label{N10EX}
\begin{center}

\begin{tabular}{|l|l|l|l|l| }
\hline
\ \ \ \ \ \ \  Method\ \ & $\ \ \ \ \ \  \nu$ & $\ \ \ \ \ \ \eta$ & $\ \ \ \ \ \omega$   \\ \hline
\ \ \begin{tabular}[c]{@{}l@{}}  \ \ \ MC(2022)\\ \ \ \  $\varepsilon^6$: (This work)\\   \ \ \  $\varepsilon^7$: (This work)\\ \ \ \ NPRG(2020)\\\end{tabular}\ \ & \begin{tabular}[c]{@{}l@{}} 0.8797(9) \\ 0.8751(18)\\ 0.8792(9 )\\ 0.8776(10)\\ \end{tabular} & \begin{tabular}[c]{@{}l@{}}0.02302(12)\\   0.02402(25)\\0.0239(20)\\ 0.0231(6)\\ \end{tabular} & \begin{tabular}[c]{@{}l@{}} 0.816(16)\\ 0.7949(83)\\ 0.7894(18)\\0.807(7)\\ \end{tabular}  \\ \hline 

\end{tabular}%
\end{center}
\end{table}
One can realize that for this order, where $N$ is relatively  large,  the precision of  our results compete with the MC and NPRG results ( sometimes our results are even more precise)
\section{Critical exponents for the \texorpdfstring{O$\left(  20\right)$}{TEXT} and \texorpdfstring{O$\left( 100\right)$}{TEXT} cases}\label{ON20100}

In this section we  aim to cover other cases for relatively large $N$. For such cases and specially for high  order terms, the coefficients in Eqs.(\ref{7Lexponenets}) are taking smaller values than the corresponding  ones for smaller $N$ values. In other words, the number of coefficients just before the smallest one  increases ( see Fig.\ref{CNvN}) which means that such series can be optimally truncated to give good results without treatment \cite{Truncation}. So when treated using summation methods, one expects to even get more accurate results. Accordingly,  it is very influential to test those cases for which such behavior is realized.

In fact, the cases $O(20)$ and  $O(100)$ have been investigated very recently using NPRG method \cite{NPR2020}.

For $O\left(  20\right)$, we obtain the $\varepsilon$-expansion for $\nu^{-1}$ to be: 

\begin{align}
\nu^{-1}  &  =2- 0.785714 \varepsilon - 0.152332 \varepsilon^2 + 0.0285453 \varepsilon^3 - 0.0935512 \varepsilon^4\nonumber\\
&   + 0.157139 \varepsilon^5 - 0.297871 \varepsilon^6 + 0.660812 \varepsilon^{7}.
\end{align}
With the coefficients of this series as input, our entire-hypergeometric approximation predicts the result $0.94784(52)$. The NPRG prediction is $0.9409(6)$ \cite{NPR2020} and the CB result is $0.9416(87)$ \cite{CB2015} while Large-N prediction gives the result $0.941(5)$ \cite{LargN-78,LNIV,LNIII}.

The the $\varepsilon$-expansion for the exponent $\eta$ can be obtained also as:

\begin{align*}
\eta &  =0.0140306 \varepsilon^2 + 0.00443825 \varepsilon^3 - 0.0047974 \varepsilon^4 + 0.00137683 \varepsilon^5 \\
& -0.00568428 \varepsilon^6 + 0.00995186 \varepsilon^{7}.
\end{align*}
Our approximation then gives the result $\eta=0.01319(38)$. The NPRG result is $0.0129(3)$  \cite{NPR2020} and CB gives the result $0.0128(16)$ \cite{CB2015} while the Large-N expansion gives the value $0.0128(2)$ \cite{LargN-78,LNIV,LNIII}.

We obtained also the $\varepsilon$-expansion for the approach to scaling exponent $\omega$ as:

\begin{align*}
\omega &  \varepsilon - 0.283163 \varepsilon^2 + 0.394011 \varepsilon^3 - 0.738584 \varepsilon^4 + 1.57377 \varepsilon^5 \\
&  - 3.58657 \varepsilon^6 + 8.96908 x^7\varepsilon^{7}.
\end{align*}
Our approximation for that case is $\omega=0.865(5)$. The NPRG results is $0.887(2)$ \cite{NPR2020} while the Large-N approximation gives the result $0.888(3)$ \cite{LargN-78,LNIV,LNIII}. We listed these results in table-\ref{N20EX}

\begin{table}[H]
\caption{\protect\scriptsize
Six- and seven-loop entire-hypergeometric resummation results for the critical exponents of the $O(20)$ vector model. For comparison, we include NPRG results \cite{NPR2020}, conformal-bootstrap (CB) estimates \cite{CB2015}, and large-$N$ approximations from Refs.~\cite{LargN-78,LNIV,LNIII}.}
\label{N20EX}
\begin{center}

\begin{tabular}{|l|l|l|l|l| }
\hline
\ \ \ \ \ \ \  Method\ \ & $\ \ \ \ \ \  \nu$ & $\ \ \ \ \ \ \eta$ & $\ \ \ \ \ \omega$   \\ \hline
\ \ \begin{tabular}[c]{@{}l@{}}  \ \ \ $\varepsilon^7$: (This work)\\ \ \ \ $\varepsilon^6$: (This work)\\ \ \ \  NPRG(2020)\\ \ \ \ \ CB\\\  \ \ \ Large-N\\\end{tabular}\ \ & \begin{tabular}[c]{@{}l@{}} 0.94784(52)\\ 0.9483(11)\\ 0.9409(6)\\ 0.9416(87)\\ 0.941(5)\\ \end{tabular} & \begin{tabular}[c]{@{}l@{}}0.01319(38)\\ 0.0138(8)\\  0.0129(3)\\ 0.0128(16)\\ 0.0128(2)\\ \end{tabular} & \begin{tabular}[c]{@{}l@{}} 0.865(5)\\0.8646(91)\\  0.887(2)\\ \ \ ----- \\ 0.807(7)\\\end{tabular}  \\ \hline 

\end{tabular}%
\end{center}
\end{table}

The $O(100)$ have been also exposed to investigation in  the references \cite{NPR2020,LargN-78,LNIV,LNIII}. In view of the seven-loop  perturbative RG, one can obtain the $\varepsilon$-expansion for $\nu^{-1}$ as:
\begin{align}
\nu^{-1}  &  =2- 0.944444 \varepsilon - 0.0544124 \varepsilon^2 + 0.00408979 \varepsilon^3 + 0.00164863 \varepsilon^4 \nonumber\\
& +  0.00924964 \varepsilon^5 - 0.00857316  \varepsilon^6 + 0.00876677 \varepsilon^{7},
\end{align}
Similarly, from the recent seven-loop $g$-expansion \cite{Schnetz2018,Schnetz2023,Schnetz-06-23}, we obtained  the seven-loop $\varepsilon$-expansion for the exponents $\eta$ and $\omega$ as:

\begin{align*}
\eta &  =0.00437243 \varepsilon^2 - 0.000386861 \varepsilon^3 - 0.00149233 \varepsilon^4 +  0.0000333565 \varepsilon^5  \\
& + 0.0000795974  \varepsilon^6 + 0.000139676 \varepsilon^{7},
\end{align*}

\begin{align*}
\omega &= \varepsilon-0.0807613 \varepsilon^2 + 0.0832863 \varepsilon^3 - 0.0550293 \varepsilon^4 + 0.0756902 \varepsilon^5\\
& - 0.0925964  \varepsilon^6 + 0.11074\varepsilon^{7}.
\end{align*}
 With these information as input, our entire-hypergeometric accurate predictions  for the exponents $\nu,\eta$ and $\omega$ are listed in table-\ref{N100EX} and compared to predictions from other methods.

\begin{table}[H]
\caption{\protect\scriptsize
Six- and seven-loop entire-hypergeometric resummation results for the critical exponents of the $O(100)$ vector model. For comparison, we also include NPRG results \cite{NPR2020} and large-$N$ approximations from Refs.~\cite{LargN-78,LNIV,LNIII}.
}
\label{N100EX}
\begin{center}

\begin{tabular}{|l|l|l|l|l| }
\hline
\ \ \ \ \ \ \  Method\ \ & $\ \ \ \ \ \  \nu$ & $\ \ \ \ \ \ \eta$ & $\ \ \ \ \ \omega$   \\ \hline
\ \ \begin{tabular}[c]{@{}l@{}}  \ \ \ $\varepsilon^7$: (This work)\\ \ \ \  $\varepsilon^6$: (This work)\\ \ \ \  NPRG(2020)\\   \ \ \ Large-N\\\end{tabular}\ \ & \begin{tabular}[c]{@{}l@{}} 0.9885(2) \\0.9894(41)\\0.9888(2)\\0.9890(2)\\  \end{tabular} & \begin{tabular}[c]{@{}l@{}}0.00273(13)\\ 0.002519(5)\\  0.00268(4)\\ 0.002681(1)\\  \end{tabular} & \begin{tabular}[c]{@{}l@{}} 0.98112(55)\\ 0.9817(36)\\ 0.9770(8)\\ 0.9782(2)\\\end{tabular}  \\ \hline 

\end{tabular}%
\end{center}
\end{table}

\section{Summary and conclusions}\label{conc}
In recent years, the critical phenomena within the $O(N)$-vector model has been investigated using important but sophisticated techniques \cite{MCON4,NPR2020,CB2016}. On the other hand the seven-loop perturbative renormalization group results for different amplitudes have been obtained recently too \cite{Schnetz2023,Schnetz2018,Schnetz-06-23}. At the critical point, one can obtain more efficient perturbation series in terms of the parameter $\varepsilon=D-4$. The series is divergent but asymptotic and thus it should be followed by an approximation algorithm (resummation, say). We realized that, for large $N$, other than the first term in the series, the coefficients in the series are monotonic decreasing functions in $N$. The first term however, is increasing as $N$  increases.  Such behavior leads to a shift of the smallest coefficient toward higher terms for larger $N$. In other words, the optimal truncation of the series will have more terms as $N$ increases which leads to better results drawn from the truncated series without any treatment \cite{Truncation}. Such realization leads us to conclude that when such series is followed by resummation algorithm, the precision  of the results will be better than those obtained for small $N$ cases. 

Ironed by the expectations mentioned above, we calculated the critical exponents for the  $O(N)$-vector model using our entire-hypergeometric resummation algorithm \cite{entire} but for $N \ge 4$.  As it was expected, we obtained precise results for the critical exponents $\nu,\eta$ and $\omega$. Aligned with the expectations above that for larger $N$ we get more precise results, our results show such expectations  very clearly. The precision is very competitive to more sophisticated tools like Monte Carlo simulation and conformal bootstrap calculations.  To have an idea about the importance of our results, we mention that for $N=4$ (for instance), the Monte Carlo calculations elapsed $8.5$ years of CPU time. Our algorithm is in fact simple, fast as well as leading to precise results for the cases under consideration.

For small N, our earlier study \cite{lambda} demonstrated that the seven-loop resummation of the $O(2)$ critical exponent $\nu$ suffers from uncertainties nearly an order of magnitude larger than those of experiment, Monte-Carlo simulations, and conformal bootstrap. Consequently, the seven loop perturbative RG could not play any meaningful role in the relatively long-standing $\lambda$-point puzzle of the ${}^4He$  superfluid transition. Remarkably, this situation changes at large N: the same seven-loop resummation now delivers predictions with  precision  that rivals those  of the most advanced Monte-Carlo and nonperturbative RG approaches.

\section*{Declaration of generative AI and AI-assisted technologies in the writing process} 
During the preparation of this work the author(s) used [CHATGPT] in order to polish the writeup of only the Abstract, the first paragraph in the introduction section and the last paragraph of this work. Also, some scattered lines have been polished  using the same tool. After using this tool/service, the author(s) reviewed and edited the content as needed and take(s) full responsibility for the content of the publication.
\bibliography{RGLareN}

\begin{thebibliography}{22}%
\makeatletter
\providecommand \@ifxundefined [1]{%
 \@ifx{#1\undefined}
}%
\providecommand \@ifnum [1]{%
 \ifnum #1\expandafter \@firstoftwo
 \else \expandafter \@secondoftwo
 \fi
}%
\providecommand \@ifx [1]{%
 \ifx #1\expandafter \@firstoftwo
 \else \expandafter \@secondoftwo
 \fi
}%
\providecommand \natexlab [1]{#1}%
\providecommand \enquote  [1]{``#1''}%
\providecommand \bibnamefont  [1]{#1}%
\providecommand \bibfnamefont [1]{#1}%
\providecommand \citenamefont [1]{#1}%
\providecommand \href@noop [0]{\@secondoftwo}%
\providecommand \href [0]{\begingroup \@sanitize@url \@href}%
\providecommand \@href[1]{\@@startlink{#1}\@@href}%
\providecommand \@@href[1]{\endgroup#1\@@endlink}%
\providecommand \@sanitize@url [0]{\catcode `\\12\catcode `\$12\catcode
  `\&12\catcode `\#12\catcode `\^12\catcode `\_12\catcode `\%12\relax}%
\providecommand \@@startlink[1]{}%
\providecommand \@@endlink[0]{}%
\providecommand \url  [0]{\begingroup\@sanitize@url \@url }%
\providecommand \@url [1]{\endgroup\@href {#1}{\urlprefix }}%
\providecommand \urlprefix  [0]{URL }%
\providecommand \Eprint [0]{\href }%
\providecommand \doibase [0]{https://doi.org/}%
\providecommand \selectlanguage [0]{\@gobble}%
\providecommand \bibinfo  [0]{\@secondoftwo}%
\providecommand \bibfield  [0]{\@secondoftwo}%
\providecommand \translation [1]{[#1]}%
\providecommand \BibitemOpen [0]{}%
\providecommand \bibitemStop [0]{}%
\providecommand \bibitemNoStop [0]{.\EOS\space}%
\providecommand \EOS [0]{\spacefactor3000\relax}%
\providecommand \BibitemShut  [1]{\csname bibitem#1\endcsname}%
\let\auto@bib@innerbib\@empty
%</preamble>
\bibitem [{\citenamefont {Pisarski}\ and\ \citenamefont {Wilczek}(1984)}]{QCD}%
  \BibitemOpen
  \bibfield  {author} {\bibinfo {author} {\bibfnamefont {R.~D.}\ \bibnamefont
  {Pisarski}}\ and\ \bibinfo {author} {\bibfnamefont {F.}~\bibnamefont
  {Wilczek}},\ }\bibfield  {title} {\bibinfo {title} {{Remarks on the chiral
  phase transition in chromodynamics}},\ }\href
  {https://doi.org/10.1103/PhysRevD.29.338} {\bibfield  {journal} {\bibinfo
  {journal} {Phys. Rev. D}\ }\textbf {\bibinfo {volume} {29}},\ \bibinfo
  {pages} {338} (\bibinfo {year} {1984})}\BibitemShut {NoStop}%
\bibitem [{\citenamefont {Hasenbusch}(2022)}]{MCON4}%
  \BibitemOpen
  \bibfield  {author} {\bibinfo {author} {\bibfnamefont {M.}~\bibnamefont
  {Hasenbusch}},\ }\bibfield  {title} {\bibinfo {title} {{Three-dimensional
  $O(N)$-invariant $\phi^4$ models at criticality for $n\ge 4$}},\ }\href
  {https://doi.org/10.1103/PhysRevB.105.054428} {\bibfield  {journal} {\bibinfo
   {journal} {Phys. Rev. B}\ }\textbf {\bibinfo {volume} {105}},\ \bibinfo
  {pages} {054428} (\bibinfo {year} {2022})}\BibitemShut {NoStop}%
\bibitem [{\citenamefont {{De Polsi}}\ \emph {et~al.}(2020)\citenamefont {{De
  Polsi}}, \citenamefont {Balog}, \citenamefont {Tissier},\ and\ \citenamefont
  {Wschebor}}]{NPR2020}%
  \BibitemOpen
  \bibfield  {author} {\bibinfo {author} {\bibfnamefont {G.}~\bibnamefont {{De
  Polsi}}}, \bibinfo {author} {\bibfnamefont {I.}~\bibnamefont {Balog}},
  \bibinfo {author} {\bibfnamefont {M.}~\bibnamefont {Tissier}},\ and\ \bibinfo
  {author} {\bibfnamefont {N.}~\bibnamefont {Wschebor}},\ }\bibfield  {title}
  {\bibinfo {title} {{Precision calculation of critical exponents in the O(N)
  universality classes with the nonperturbative renormalization group}},\
  }\href {https://doi.org/10.1103/PhysRevE.101.042113} {\bibfield  {journal}
  {\bibinfo  {journal} {Phys. Rev. E}\ }\textbf {\bibinfo {volume} {101}},\
  \bibinfo {pages} {042113} (\bibinfo {year} {2020})}\BibitemShut {NoStop}%
\bibitem [{\citenamefont {Carrington}\ \emph {et~al.}(2018)\citenamefont
  {Carrington}, \citenamefont {Friesen}, \citenamefont {Meggison},
  \citenamefont {Phillips}, \citenamefont {Pickering},\ and\ \citenamefont
  {Sohrabi}}]{2PILN}%
  \BibitemOpen
  \bibfield  {author} {\bibinfo {author} {\bibfnamefont {M.~E.}\ \bibnamefont
  {Carrington}}, \bibinfo {author} {\bibfnamefont {S.~A.}\ \bibnamefont
  {Friesen}}, \bibinfo {author} {\bibfnamefont {B.~A.}\ \bibnamefont
  {Meggison}}, \bibinfo {author} {\bibfnamefont {C.~D.}\ \bibnamefont
  {Phillips}}, \bibinfo {author} {\bibfnamefont {D.}~\bibnamefont
  {Pickering}},\ and\ \bibinfo {author} {\bibfnamefont {K.}~\bibnamefont
  {Sohrabi}},\ }\bibfield  {title} {\bibinfo {title} {{2PI effective theory at
  next-to-leading order using the functional renormalization group}},\ }\href
  {https://doi.org/10.1103/PhysRevD.97.036005} {\bibfield  {journal} {\bibinfo
  {journal} {Phys. Rev. D}\ }\textbf {\bibinfo {volume} {97}},\ \bibinfo
  {pages} {036005} (\bibinfo {year} {2018})}\BibitemShut {NoStop}%
\bibitem [{\citenamefont {Saito}\ \emph {et~al.}(2012)\citenamefont {Saito},
  \citenamefont {Fujii}, \citenamefont {Itakura},\ and\ \citenamefont
  {Morimatsu}}]{2PI2011}%
  \BibitemOpen
  \bibfield  {author} {\bibinfo {author} {\bibfnamefont {Y.}~\bibnamefont
  {Saito}}, \bibinfo {author} {\bibfnamefont {H.}~\bibnamefont {Fujii}},
  \bibinfo {author} {\bibfnamefont {K.}~\bibnamefont {Itakura}},\ and\ \bibinfo
  {author} {\bibfnamefont {O.}~\bibnamefont {Morimatsu}},\ }\bibfield  {title}
  {\bibinfo {title} {{Critical exponents from the two-particle irreducible
  $1/N$ expansion}},\ }\href {https://doi.org/10.1103/PhysRevD.85.065019}
  {\bibfield  {journal} {\bibinfo  {journal} {Phys. Rev. D}\ }\textbf {\bibinfo
  {volume} {85}},\ \bibinfo {pages} {065019} (\bibinfo {year}
  {2012})}\BibitemShut {NoStop}%
\bibitem [{\citenamefont {Schnetz}(2023{\natexlab{a}})}]{Schnetz2023}%
  \BibitemOpen
  \bibfield  {author} {\bibinfo {author} {\bibfnamefont {O.}~\bibnamefont
  {Schnetz}},\ }\bibfield  {title} {\bibinfo {title} {{$\phi^4$ Theory At Seven
  Loops}},\ }\href {https://doi.org/10.1103/PhysRevD.107.036002} {\bibfield
  {journal} {\bibinfo  {journal} {Phys. Rev. D}\ }\textbf {\bibinfo {volume}
  {107}},\ \bibinfo {pages} {036002} (\bibinfo {year}
  {2023}{\natexlab{a}})}\BibitemShut {NoStop}%
\bibitem [{\citenamefont {Schnetz}(2023{\natexlab{b}})}]{Schnetz-06-23}%
  \BibitemOpen
  \bibfield  {author} {\bibinfo {author} {\bibfnamefont {O.}~\bibnamefont
  {Schnetz}},\ }\bibfield  {title} {\bibinfo {title} {{Maple package
  HyperlogProcedrues}},\ }\href
  {https://www.math.fau.de/person/oliver-schnetz/} {\  (\bibinfo {year}
  {2023}{\natexlab{b}})}\BibitemShut {NoStop}%
\bibitem [{\citenamefont {Schnetz}(2018)}]{Schnetz2018}%
  \BibitemOpen
  \bibfield  {author} {\bibinfo {author} {\bibfnamefont {O.}~\bibnamefont
  {Schnetz}},\ }\bibfield  {title} {\bibinfo {title} {{Numbers and functions in
  quantum field theory}},\ }\href {https://doi.org/10.1103/PhysRevD.97.085018}
  {\bibfield  {journal} {\bibinfo  {journal} {Phys. Rev. D}\ }\textbf {\bibinfo
  {volume} {97}},\ \bibinfo {pages} {085018} (\bibinfo {year} {2018})},\
  \Eprint {https://arxiv.org/abs/1606.08598} {arXiv:1606.08598} \BibitemShut
  {NoStop}%
\bibitem [{\citenamefont {Kompaniets}\ and\ \citenamefont
  {Panzer}(2017)}]{Borel-6L}%
  \BibitemOpen
  \bibfield  {author} {\bibinfo {author} {\bibfnamefont {M.~V.}\ \bibnamefont
  {Kompaniets}}\ and\ \bibinfo {author} {\bibfnamefont {E.}~\bibnamefont
  {Panzer}},\ }\bibfield  {title} {\bibinfo {title} {{Minimally subtracted
  six-loop renormalization of $O(n)$-symmetric $\phi^4$ theory and critical
  exponents}},\ }\href {https://doi.org/10.1103/PhysRevD.96.036016} {\bibfield
  {journal} {\bibinfo  {journal} {Phys. Rev. D}\ }\textbf {\bibinfo {volume}
  {96}},\ \bibinfo {pages} {036016} (\bibinfo {year} {2017})}\BibitemShut
  {NoStop}%
\bibitem [{\citenamefont {Elkamash}\ \emph {et~al.}(2023)\citenamefont
  {Elkamash}, \citenamefont {Abdelhamid},\ and\ \citenamefont
  {Shalaby}}]{entire}%
  \BibitemOpen
  \bibfield  {author} {\bibinfo {author} {\bibfnamefont {I.}~\bibnamefont
  {Elkamash}}, \bibinfo {author} {\bibfnamefont {H.~M.}\ \bibnamefont
  {Abdelhamid}},\ and\ \bibinfo {author} {\bibfnamefont {A.~M.}\ \bibnamefont
  {Shalaby}},\ }\bibfield  {title} {\bibinfo {title} {{Entire hypergeometric
  approximants for the ground state energy perturbation series of the quartic,
  sextic and octic anharmonic oscillators}},\ }\href
  {https://doi.org/10.1016/j.aop.2023.169427} {\bibfield  {journal} {\bibinfo
  {journal} {Ann. Phys. (N. Y).}\ }\textbf {\bibinfo {volume} {457}},\ \bibinfo
  {pages} {169427} (\bibinfo {year} {2023})}\BibitemShut {NoStop}%
\bibitem [{\citenamefont {Dunne}\ and\ \citenamefont
  {Meynig}(2022)}]{Renormalon22}%
  \BibitemOpen
  \bibfield  {author} {\bibinfo {author} {\bibfnamefont {G.~V.}\ \bibnamefont
  {Dunne}}\ and\ \bibinfo {author} {\bibfnamefont {M.}~\bibnamefont {Meynig}},\
  }\bibfield  {title} {\bibinfo {title} {{Instantons or renormalons? Remarks on
  theory in the MS scheme}},\ }\href
  {https://doi.org/10.1103/PhysRevD.105.025019} {\bibfield  {journal} {\bibinfo
   {journal} {Phys. Rev. D}\ }\textbf {\bibinfo {volume} {105}},\ \bibinfo
  {pages} {025019} (\bibinfo {year} {2022})}\BibitemShut {NoStop}%
\bibitem [{\citenamefont {Suslov}(2005)}]{Ren-2005}%
  \BibitemOpen
  \bibfield  {author} {\bibinfo {author} {\bibfnamefont {I.~M.}\ \bibnamefont
  {Suslov}},\ }\bibfield  {title} {\bibinfo {title} {{Divergent perturbation
  series}},\ }\href {https://doi.org/10.1134/1.1995802} {\bibfield  {journal}
  {\bibinfo  {journal} {J. Exp. Theor. Phys.}\ }\textbf {\bibinfo {volume}
  {100}},\ \bibinfo {pages} {1188} (\bibinfo {year} {2005})}\BibitemShut
  {NoStop}%
\bibitem [{\citenamefont {Beneke}(1999)}]{ren-rep}%
  \BibitemOpen
  \bibfield  {author} {\bibinfo {author} {\bibfnamefont {M.}~\bibnamefont
  {Beneke}},\ }\bibfield  {title} {\bibinfo {title} {{Renormalons}},\ }\href
  {https://doi.org/10.1016/S0370-1573(98)00130-6} {\bibfield  {journal}
  {\bibinfo  {journal} {Phys. Rep.}\ }\textbf {\bibinfo {volume} {317}},\
  \bibinfo {pages} {1} (\bibinfo {year} {1999})}\BibitemShut {NoStop}%
\bibitem [{\citenamefont {Kilbas}\ \emph {et~al.}(2016)\citenamefont {Kilbas},
  \citenamefont {Saxena}, \citenamefont {Saigo},\ and\ \citenamefont
  {Trujillo}}]{Analytic2016}%
  \BibitemOpen
  \bibfield  {author} {\bibinfo {author} {\bibfnamefont {A.~A.}\ \bibnamefont
  {Kilbas}}, \bibinfo {author} {\bibfnamefont {R.~K.}\ \bibnamefont {Saxena}},
  \bibinfo {author} {\bibfnamefont {M.}~\bibnamefont {Saigo}},\ and\ \bibinfo
  {author} {\bibfnamefont {J.~J.}\ \bibnamefont {Trujillo}},\ }\bibfield
  {title} {\bibinfo {title} {{The generalized hypergeometric function as the
  Meijer G-function}},\ }\href {https://doi.org/10.1515/anly-2015-5001}
  {\bibfield  {journal} {\bibinfo  {journal} {Analysis}\ }\textbf {\bibinfo
  {volume} {36}},\ \bibinfo {pages} {1} (\bibinfo {year} {2016})}\BibitemShut
  {NoStop}%
\bibitem [{\citenamefont {Shalaby}(2020)}]{lambda}%
  \BibitemOpen
  \bibfield  {author} {\bibinfo {author} {\bibfnamefont {A.~M.}\ \bibnamefont
  {Shalaby}},\ }\bibfield  {title} {\bibinfo {title} {{$\lambda$-point anomaly
  in view of the seven-loop hypergeometric resummation for the critical
  exponent $\nu$ of the $O(2) \phi^4$ model}},\ }\href
  {https://doi.org/10.1103/PhysRevD.102.105017} {\bibfield  {journal} {\bibinfo
   {journal} {Phys. Rev. D}\ }\textbf {\bibinfo {volume} {102}},\ \bibinfo
  {pages} {105017} (\bibinfo {year} {2020})}\BibitemShut {NoStop}%
\bibitem [{\citenamefont {Kleinert}\ and\ \citenamefont
  {Schulte-Frohlinde}(2001)}]{Kleinert-Borel}%
  \BibitemOpen
  \bibfield  {author} {\bibinfo {author} {\bibfnamefont {H.}~\bibnamefont
  {Kleinert}}\ and\ \bibinfo {author} {\bibfnamefont {V.}~\bibnamefont
  {Schulte-Frohlinde}},\ }\href {https://doi.org/10.1142/4733} {\emph {\bibinfo
  {title} {{Critical Properties of {$\phi^4$} -Theories}}}}\ (\bibinfo
  {publisher} {WORLD SCIENTIFIC},\ \bibinfo {year} {2001})\BibitemShut
  {NoStop}%
\bibitem [{\citenamefont {Shalaby}(2025)}]{Truncation}%
  \BibitemOpen
  \bibfield  {author} {\bibinfo {author} {\bibfnamefont {A.~M.}\ \bibnamefont
  {Shalaby}},\ }\bibfield  {title} {\bibinfo {title} {{Accurate critical
  exponents from the optimal truncation of the $\varepsilon $-expansion within
  the $O(N)$-symmetric field theory for large $N$}},\ }\href
  {https://doi.org/10.1140/epjc/s10052-025-14473-7} {\bibfield  {journal}
  {\bibinfo  {journal} {Eur. Phys. J. C}\ }\textbf {\bibinfo {volume} {85}},\
  \bibinfo {pages} {751} (\bibinfo {year} {2025})}\BibitemShut {NoStop}%
\bibitem [{\citenamefont {Echeverri}\ \emph {et~al.}(2016)\citenamefont
  {Echeverri}, \citenamefont {von Harling},\ and\ \citenamefont
  {Serone}}]{CB2016}%
  \BibitemOpen
  \bibfield  {author} {\bibinfo {author} {\bibfnamefont {A.~C.}\ \bibnamefont
  {Echeverri}}, \bibinfo {author} {\bibfnamefont {B.}~\bibnamefont {von
  Harling}},\ and\ \bibinfo {author} {\bibfnamefont {M.}~\bibnamefont
  {Serone}},\ }\bibfield  {title} {\bibinfo {title} {{The effective
  bootstrap}},\ }\href {https://doi.org/10.1007/JHEP09(2016)097} {\bibfield
  {journal} {\bibinfo  {journal} {J. High Energy Phys.}\ }\textbf {\bibinfo
  {volume} {2016}}\bibinfo  {number} { (9)},\ \bibinfo {pages}
  {97}}\BibitemShut {NoStop}%
\bibitem [{\citenamefont {Kos}\ \emph {et~al.}(2015)\citenamefont {Kos},
  \citenamefont {Poland}, \citenamefont {Simmons-Duffin},\ and\ \citenamefont
  {Vichi}}]{CB2015}%
  \BibitemOpen
\bibfield  {number} {  }\bibfield  {author} {\bibinfo {author} {\bibfnamefont
  {F.}~\bibnamefont {Kos}}, \bibinfo {author} {\bibfnamefont {D.}~\bibnamefont
  {Poland}}, \bibinfo {author} {\bibfnamefont {D.}~\bibnamefont
  {Simmons-Duffin}},\ and\ \bibinfo {author} {\bibfnamefont {A.}~\bibnamefont
  {Vichi}},\ }\bibfield  {title} {\bibinfo {title} {{Bootstrapping the O(N)
  archipelago}},\ }\href {https://doi.org/10.1007/JHEP11(2015)106} {\bibfield
  {journal} {\bibinfo  {journal} {J. High Energy Phys.}\ }\textbf {\bibinfo
  {volume} {2015}}\bibinfo  {number} { (11)},\ \bibinfo {pages}
  {106}}\BibitemShut {NoStop}%
\bibitem [{\citenamefont {Okabe}\ and\ \citenamefont {Oku}(1978)}]{LargN-78}%
  \BibitemOpen
\bibfield  {number} {  }\bibfield  {author} {\bibinfo {author} {\bibfnamefont
  {Y.}~\bibnamefont {Okabe}}\ and\ \bibinfo {author} {\bibfnamefont
  {M.}~\bibnamefont {Oku}},\ }\bibfield  {title} {\bibinfo {title} {{1/n
  Expansion Up to Order 1/n2. III: Critical Exponents and for d=3}},\ }\href
  {https://doi.org/10.1143/PTP.60.1287} {\bibfield  {journal} {\bibinfo
  {journal} {Prog. Theor. Phys.}\ }\textbf {\bibinfo {volume} {60}},\ \bibinfo
  {pages} {1287} (\bibinfo {year} {1978})}\BibitemShut {NoStop}%
\bibitem [{\citenamefont {Vasil'ev}\ \emph {et~al.}(1982)\citenamefont
  {Vasil'ev}, \citenamefont {Pis'mak},\ and\ \citenamefont {Khonkonen}}]{LNIV}%
  \BibitemOpen
  \bibfield  {author} {\bibinfo {author} {\bibfnamefont {A.~N.}\ \bibnamefont
  {Vasil'ev}}, \bibinfo {author} {\bibfnamefont {Y.~M.}\ \bibnamefont
  {Pis'mak}},\ and\ \bibinfo {author} {\bibfnamefont {Y.~R.}\ \bibnamefont
  {Khonkonen}},\ }\bibfield  {title} {\bibinfo {title} {{$1/n$ Expansion:
  Calculation of the exponent $\nu$ in the order $1/{n^3}$ by the Conformal
  Bootstrap Method}},\ }\href {https://doi.org/10.1007/BF01015292} {\bibfield
  {journal} {\bibinfo  {journal} {Theor. Math. Phys.}\ }\textbf {\bibinfo
  {volume} {50}},\ \bibinfo {pages} {127} (\bibinfo {year} {1982})}\BibitemShut
  {NoStop}%
\bibitem [{\citenamefont {Broadhurst}\ \emph {et~al.}(1996)\citenamefont
  {Broadhurst}, \citenamefont {Gracey},\ and\ \citenamefont {Kreimer}}]{LNIII}%
  \BibitemOpen
  \bibfield  {author} {\bibinfo {author} {\bibfnamefont {D.~J.}\ \bibnamefont
  {Broadhurst}}, \bibinfo {author} {\bibfnamefont {J.~A.}\ \bibnamefont
  {Gracey}},\ and\ \bibinfo {author} {\bibfnamefont {D.}~\bibnamefont
  {Kreimer}},\ }\bibfield  {title} {\bibinfo {title} {{Beyond the triangle and
  uniqueness relations: non-zeta counterterms at large N from positive
  knots}},\ }\href {http://arxiv.org/abs/hep-th/9607174} {\bibfield  {journal}
  {\bibinfo  {journal} {Z.Phys. C75 559-574}\ } (\bibinfo {year} {1996})},\
  \Eprint {https://arxiv.org/abs/9607174} {arXiv:9607174 [hep-th]} \BibitemShut
  {NoStop}%
\end{thebibliography}%
\end{document}